\documentclass[aps,prb,reprint,superscriptaddress,amsmath,amssymb,longbibliography,floatfix]{revtex4-2}
\usepackage[T1]{fontenc}
\usepackage[utf8]{inputenc}
\usepackage{newtxtext,newtxmath,microtype,graphicx,bm,mathtools,placeins,xcolor,booktabs}
\usepackage{tikz}
\usetikzlibrary{arrows.meta}
\definecolor{modelInk}{HTML}{252A32}
\definecolor{modelBond}{HTML}{606874}
\definecolor{modelBlue}{HTML}{2458C6}
\definecolor{modelNormal}{HTML}{F7F8FA}
\definecolor{modelSuper}{HTML}{EAF0FC}
\usepackage[colorlinks=true,allcolors=blue!45!black]{hyperref}
\graphicspath{{figures/}}
\newcommand{\ii}{\mathrm{i}}
\newcommand{\Id}{\mathbb I}

\begin{document}
\title{Andreev Reflection at a Graphene Normal–Superconductor Interface: A Lattice–Continuum Comparison}

\author{Dar Faheem Ul Haq}
\affiliation{Department of Physics, Aligarh Muslim University, Aligarh 202002, India}

\author{S. S. Z. Ashraf}
\affiliation{Department of Physics, Aligarh Muslim University, Aligarh 202002, India}

\date{26 October 2026}
\begin{abstract}
We compare a quadratic continuum model with a nearest-neighbour tight-binding model to study Andreev reflection at a clean graphene normal–superconductor interface. For abrupt changes in the chemical potential and pairing potential, with the hopping amplitude kept unchanged, the lattice model leads to four independent boundary conditions. The reflection probabilities obtained from this approach agree with those calculated independently using Kwant, to numerical precision for the channels examined.

We also investigate trigonal warping, the threefold anisotropy of the graphene band structure, which affects the relationship between momentum and group velocity. For \((\mu_N,\mu_S,\Delta_0,E)/t_0=(0.5,1,0.1,0.05)\), the maximum in the Andreev-reflection probability for the \(\xi=+1\) valley occurs at approximately \(19.9^\circ\) in the lattice model, compared with \(3.4^\circ\) in the quadratic continuum model. However, these parameters lie outside the regime in which the low-energy continuum approximation is expected to be reliable. When the chemical potentials, superconducting gap, and excitation energy are reduced together while their ratios are kept fixed, the angular response near the central direction and the conductance become more consistent with the lattice results.

A further complication is that the quadratic continuum boundary problem requires short-wavelength auxiliary modes that lie outside the range of validity of the valley expansion. The differences between the two models therefore cannot be explained by bulk band dispersion alone. Our results highlight that a small excitation energy relative to the chemical potential does not, by itself, guarantee the validity of a continuum description. A reliable assessment must also consider the relevant energy scales and modes in both the normal and superconducting regions, as well as the assumptions used to match the wavefunctions at the interface.
\end{abstract}
\maketitle

\section{Introduction}
Andreev reflection couples normal electron transport to a superconducting
condensate: an incident electron is reflected as a hole and a Cooper
pair enters the superconductor
\cite{Andreev1964,BlonderTinkhamKlapwijk1982}. In graphene, this process
depends on whether the reflected hole belongs to the same band as the
electron or to the opposite band. The Dirac description predicts a
crossover between retro and specular reflection as doping and
excitation energy are varied \cite{Beenakker2006,Beenakker2008}.
The normal state already shows how strongly graphene transport depends
on incident direction. Klein transmission through an electrostatic
barrier \cite{Katsnelson2006} and angular filtering by a smooth
P--N junction \cite{CheianovFalko2006} follow from the structure of
the Dirac states. Interference measurements in graphene
heterojunctions have resolved signatures of this chiral transmission
\cite{YoungKim2009}. At an NS interface the reflected hole must also
be matched to superconducting electron--hole superpositions.
Its relation to interband transmission connects the superconducting
problem to Klein tunnelling in bipolar graphene
\cite{BeenakkerAkhmerovRecherTworzydlo2008}.

Graphene provides direct control of carrier density
\cite{Novoselov2004}, and its quantum Hall response established the
Dirac character of low-energy carriers \cite{Novoselov2005,Zhang2005}.
Away from the valleys, the honeycomb dispersion becomes nonlinear
and its nearly circular constant-energy contours develop a threefold
distortion, known as trigonal warping \cite{Wallace1947,CastroNeto2009}.
The linear Dirac approximation gives an isotropic dispersion near each valley, whereas a second-order expansion of the lattice Hamiltonian captures the leading trigonal-warping correction. This correction modifies both the constant-energy contours and the directions of the group velocities, thereby affecting the sublattice spinors and boundary matching at the interface. A small excitation energy \(E\) does not necessarily imply that the relevant states lie close to a Dirac point in both regions. Their positions within the graphene band structure also depend on the local chemical potentials, particularly in the superconducting region, where strong doping can take the relevant states outside the range of validity of a low-energy valley expansion. In particular, a strongly doped superconducting
region can sample a part of the band for which the linear theory is
inaccurate.

\begin{figure}[tbp]
\centering
\definecolor{figInk}{HTML}{111111}
\definecolor{figBond}{HTML}{1F2937}
\definecolor{figNormal}{HTML}{F1F3F5}
\definecolor{figSuper}{HTML}{F2B84B}
\definecolor{vecRed}{HTML}{C8102E}
\definecolor{zigHighlight}{HTML}{7B1FA2}
% Self-contained vector artwork; all lattice coordinates follow the text.
\resizebox{\columnwidth}{!}{%
\begin{tikzpicture}[x=1cm,y=1cm,
  font=\sffamily\fontsize{9}{11}\selectfont,text=figInk,
  line cap=round,line join=round,
  vector/.style={draw=vecRed,line width=2.6pt,
    -{Stealth[length=3.4mm,width=2.8mm]},shorten >=4pt},
  atom A/.style={draw=figInk,fill=figInk,line width=.9pt},
  atom B/.style={draw=figInk,fill=white,line width=1.7pt}]
  \path[use as bounding box] (0,-.15) rectangle (8.6,5.95);

  % (a) honeycomb lattice across the NS interface
  \node[font=\sffamily\bfseries\fontsize{12}{13}\selectfont] at (.18,5.72) {a};
  \node[font=\sffamily\bfseries\fontsize{9.5}{11}\selectfont] at (2.25,5.72) {Normal graphene};
  \node[font=\sffamily\bfseries\fontsize{9.5}{11}\selectfont] at (6.55,5.72) {Proximitised graphene};
  \node at (2.25,5.34) {$\mu_N,\quad\Delta=0$};
  \node at (6.55,5.34) {$\mu_S,\quad\Delta=\Delta_0$};
  \begin{scope}
    \clip (.25,2.45) rectangle (8.35,5.10);
    \fill[figNormal] (.25,2.45) rectangle (4.40,5.10);
    \fill[figSuper] (4.40,2.45) rectangle (8.35,5.10);
    \foreach \m in {-5,...,5}{
      \foreach \n in {-5,...,5}{
        \pgfmathsetmacro{\ax}{4.40+.68*(1.5*\m-1)}
        \pgfmathsetmacro{\ay}{3.85+.68*sqrt(3)*(\n-.5*\m)}
        \foreach \sign in {-1,1}{
          \draw[figBond,line width=2.1pt]
            (\ax,\ay) -- ({\ax+.34},{\ay+\sign*.34*sqrt(3)});
        }
        \draw[figBond,line width=2.1pt] (\ax,\ay) -- ({\ax-.68},\ay);
      }
    }
    % Existing zigzag chain at m=-2; every highlighted bond still has hopping -t_0.
    \draw[zigHighlight,line width=3.3pt]
      (1.68,{3.85-.68*sqrt(3)}) -- (2.02,{3.85-.34*sqrt(3)})
      -- (1.68,3.85) -- (2.02,{3.85+.34*sqrt(3)})
      -- (1.68,{3.85+.68*sqrt(3)});
    \node[text=zigHighlight,align=center,inner sep=0pt,
      font=\sffamily\bfseries\fontsize{7.5}{8.5}\selectfont]
      at (1.30,{3.85+.34*sqrt(3)}) {Zigzag\\chain};
    \draw[figInk,line width=1.8pt,dash pattern=on 4pt off 3pt]
      (4.40,2.45) -- (4.40,5.10);
    \foreach \m in {-5,...,5}{
      \foreach \n in {-5,...,5}{
        \pgfmathsetmacro{\ax}{4.40+.68*(1.5*\m-1)}
        \pgfmathsetmacro{\ay}{3.85+.68*sqrt(3)*(\n-.5*\m)}
        \path[atom A] (\ax,\ay) circle[radius=3.1pt];
        \path[atom B] ({\ax+.34},{\ay-.34*sqrt(3)}) circle[radius=3.1pt];
      }
    }
  \end{scope}
  % panel frame
  \draw[figInk,line width=1.2pt] (.25,2.45) rectangle (8.35,5.10);

  % primitive vectors (red)
  \coordinate (O) at (2.70,{3.85-.34*sqrt(3)});
  \coordinate (P) at (3.72,{3.85-.68*sqrt(3)});
  \coordinate (Q) at (2.70,{3.85+.34*sqrt(3)});
  \draw[white,line width=6pt] (O) -- (P);
  \draw[white,line width=6pt] (O) -- (Q);
  \path[atom A] (P) circle[radius=3.1pt];
  \path[atom A] (Q) circle[radius=3.1pt];
  \draw[vector] (O) -- (P);
  \draw[vector] (O) -- (Q);
  \fill[vecRed] (O) circle[radius=3.2pt];
  \node[fill=figNormal,inner sep=1.6pt,text=vecRed,
    font=\fontsize{12}{13}\selectfont] at (3.47,3.17) {$\bm a_1$};
  \node[fill=figNormal,inner sep=1.6pt,text=vecRed,
    font=\fontsize{12}{13}\selectfont] at (2.34,3.96) {$\bm a_2$};

  % axes and sublattice legend
  \draw[figInk,line width=1.4pt,-{Stealth[length=2mm]}]
    (.55,1.95) -- (1.12,1.95) node[right,inner sep=2pt] {$x$};
  \draw[figInk,line width=1.4pt,-{Stealth[length=2mm]}]
    (.55,1.95) -- (.55,2.28) node[left,inner sep=2pt] {$y$};
  \path[atom A] (2.02,2.12) circle[radius=3.1pt];
  \node[anchor=west,inner sep=3.5pt] at (2.02,2.12) {$A$};
  \path[atom B] (2.85,2.12) circle[radius=3.1pt];
  \node[anchor=west,inner sep=3.5pt] at (2.85,2.12) {$B$};
  \node at (6.46,2.12) {Every bond: $-t_0$};

  % (b) chain at fixed k_Y
  \node[font=\sffamily\bfseries\fontsize{12}{13}\selectfont] at (.18,1.62) {b};
  \node[anchor=west,font=\sffamily\bfseries\fontsize{9.5}{11}\selectfont] at (.49,1.62) {Chain at fixed $k_Y$};
  \node at (6.85,1.62) {$\eta=1+e^{\ii k_Y}$};
  \foreach \j/\m/\shade in {0/-1/figNormal,1/0/figNormal,2/1/figSuper,3/2/figSuper}{
    \pgfmathsetmacro{\cx}{1+2*\j}
    \path[draw=figInk,line width=1.1pt,dash pattern=on 3pt off 2pt,fill=\shade]
      ({\cx-.25},.30) rectangle ({\cx+1.05},1.16);
  }
  \draw[figInk,line width=2.6pt] (.65,.78) -- (8.15,.78);
  \node at (.40,.78) {$\cdots$};
  \node at (8.40,.78) {$\cdots$};
  \draw[figInk,line width=1.8pt,dash pattern=on 4pt off 3pt]
    (4.40,.20) -- (4.40,1.26);
  \foreach \j/\m in {0/-1,1/0,2/1,3/2}{
    \pgfmathsetmacro{\cx}{1+2*\j}
    \path[atom A] (\cx,.78) circle[radius=3.6pt];
    \path[atom B] ({\cx+.80},.78) circle[radius=3.6pt];
    \node at (\cx,1.02) {$A$};
    \node at ({\cx+.80},1.02) {$B$};
    \node at ({\cx+.40},.08) {$m=\m$};
  }
  \node[fill=figNormal,inner sep=1.2pt] at (1.40,.50) {$-\eta$};
  \node[fill=white,inner sep=1.2pt] at (4.40,.50) {$-1$};
\end{tikzpicture}
}

\caption{(a) Graphene NS step parallel to $\bm a_2$.
Filled and open circles denote the A and B sublattices.
Purple bonds highlight a zigzag chain with repeat direction $\bm a_2$.
The primitive translations are drawn between equivalent A sites:
$\bm a_1=a_{\rm CC}(3/2,-\sqrt3/2)$ and
$\bm a_2=a_{\rm CC}(0,\sqrt3)$.
The dashed line separates the last normal cell, $m=0$, from the first
superconducting cell, $m=1$. Every microscopic bond has hopping $-t_0$.
(b) Schematic chain at fixed $k_Y$; dashed boxes enclose cells.
In units of $t_0$, the intracell matrix elements are
$h_{AB}=-\eta$ and $h_{BA}=-\eta^*$, with $\eta=1+e^{\ii k_Y}$;
the intercell hopping is $t_{BA}=-1$, including the $B_0$--$A_1$
interface bond. Only the chemical potential and pair potential change
across the interface.}
\label{fig:model}
\end{figure}
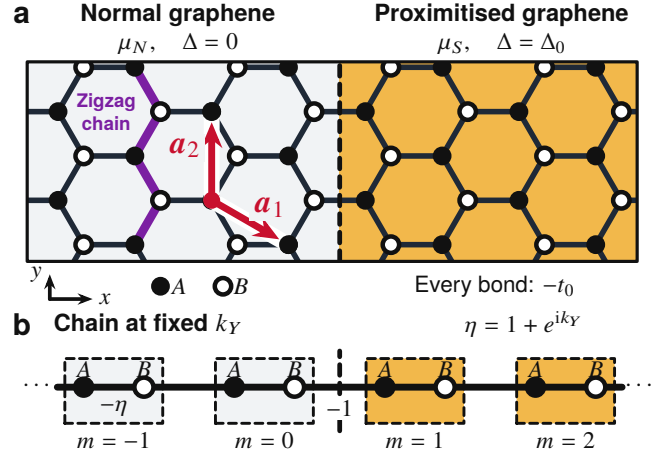

The experimental motivation extends from the Josephson effect predicted
for ballistic graphene \cite{TitovBeenakker2006} to observations of
bipolar supercurrents, multiple Andreev reflection, and tunable proximity
coupling \cite{Heersche2007,Du2008,OjedaAristizabal2009}. Ballistic
junctions resolve interference and junction-length dependence
\cite{Calado2015,BenShalom2016,Borzenets2016}; tunnelling spectroscopy
resolves Andreev states \cite{Bretheau2017}. NS conductance measurements
and spatial imaging probe the reflection process more directly
\cite{Pandey2019,Bhandari2020}. Connecting such observations to a model
requires a specified interface as well as a bulk dispersion.
Microscopic proximity models and Green-function approaches demonstrate
this dependence \cite{BursetYeyatiMartinRodero2008,Herrera2010}, while
self-consistent pairing and ribbon geometry add further structure
\cite{BlackSchafferDoniach2008,Rainis2009}. Charge transfer at metal contacts can shift graphene's local carrier
density \cite{Khomyakov2009Contacts}, so the normal and contact-side
chemical potentials need not coincide. Contact matching remains
central even at clean graphene--metal-superconductor interfaces
\cite{Bondarev2025}.

Analytical lattice mode matching provides a useful framework for testing continuum approximations. In studies of normal graphene P--N junctions, the tilted-coordinate construction has revealed oblique transmission beyond the linear regime \cite{ZhangYangChang2017,ZhangYang2018}. Here, we extend the lattice mode-matching approach to a graphene normal--superconductor (NS) interface, retaining the electron--hole structure and the decaying modes in the superconducting region. Previous continuum studies have examined the dependence of graphene NS conductance on doping, barriers, and pairing \cite{BhattacharjeeSengupta2006,LinderSudbo2008}. Building on this work, we compare a nearest-neighbour tight-binding model with quadratic and linear continuum descriptions to assess how accurately they reproduce Andreev reflection at a specified lattice interface. The chemical-potential and pairing profiles are kept fixed across the models, while the modes, currents, and boundary conditions are treated consistently within each description. We first consider a parameter regime in which the continuum approximation is expected to be unreliable, and then examine how the angular reflection response and conductance change as the relevant energy scales are reduced together at fixed ratios. This comparison helps identify the limitations of the continuum descriptions without attributing all discrepancies to trigonal warping alone. Section~II develops the lattice and continuum boundary problems. Section~III presents the angular reflection results, scaling behaviour, and conductance, followed by the conclusions in Section~IV.

\section{Lattice mode matching and continuum comparison}
\subsection{Junction geometry and tilted coordinates}

The normal region occupies $X<0$ and the proximitised region $X>0$,
with
\begin{equation}
 \mu(X)=\mu_N\Theta(-X)+\mu_S\Theta(X),\qquad
 \Delta(X)=\Delta_0\Theta(X).
 \label{eq:profiles}
\end{equation}
We consider a straight NS interface parallel to the zigzag
direction $\bm a_2$, with Cartesian $x$ normal to the interface.
The graphene lattice remains continuous across the junction.
All results below refer to this fixed interface orientation.
There is no tunnel barrier, disorder, magnetic field, or
sublattice-dependent potential. The parent superconductor enters
only through the imposed pair potential. A microscopic proximity
model would instead describe coupling to the parent material and
determine an induced spectrum \cite{McMillan1968}. Keeping the gap
prescribed isolates the present comparison from that additional
contact problem.
We set the superconducting phase to zero and consider
$0<E<\min(\mu_N,\Delta_0)$. The gap is prescribed, with no
self-consistent pairing calculation. Unless units are displayed,
energies in the Hamiltonians below are divided by the positive
hopping scale $t_0$. The plotted calculations have $\mu_N>E>0$:
the reflected hole is in the conduction band. The interband
specular regime is background motivation, rather than a regime
tested by these data. Figure~\ref{fig:model} defines the
geometry and lattice convention. The primitive translations are
\begin{equation}
 \bm a_1=a_{\rm CC}(3/2,-\sqrt3/2),\qquad
 \bm a_2=a_{\rm CC}(0,\sqrt3).
\end{equation}
Translation along $\bm a_2$ conserves $k_Y$. The Bloch phases
$k_X=\bm k\cdot\bm a_1$ and $k_Y=\bm k\cdot\bm a_2$ are tilted
coordinates. If $v_j=\partial E/\partial k_j$, the physical
velocity and incidence angle are
\begin{equation}
 \bm v=\frac{t_0}{\hbar}(\bm a_1v_X+\bm a_2v_Y),\qquad
 \theta_v=\operatorname{atan2}(v_y,v_x).
 \label{eq:angle}
\end{equation}
In components,
\begin{equation}
 v_x=\frac{t_0a_{\rm CC}}{\hbar}\frac32v_X,\qquad
 v_y=\frac{t_0a_{\rm CC}}{\hbar}\sqrt3(v_Y-v_X/2).
 \label{eq:cartvelocity}
\end{equation}
Thus $X$ labels a tilted translation, whereas Cartesian $x$ is
normal to the interface. A valley displacement
$\bm q=\bm k-\bm K_\xi$ obeys
$p_X=a_{\rm CC}(3q_x/2-\sqrt3q_y/2)$ and
$p_Y=\sqrt3a_{\rm CC}q_y$.

\begin{figure}[!t]
\centering
\definecolor{figInk}{HTML}{111111}
\definecolor{figBond}{HTML}{1F2937}
\definecolor{figNormal}{HTML}{9CC4EC}
\definecolor{figSuper}{HTML}{F2B84B}
\definecolor{vecRed}{HTML}{C8102E}
\definecolor{phaseBlue}{HTML}{0B5CAD}
\resizebox{\columnwidth}{!}{%
\begin{tikzpicture}[x=1cm,y=1cm,font=\sffamily\fontsize{9}{11}\selectfont,text=figInk,line cap=round,line join=round,
  atomA/.style={draw=figInk,fill=figInk,line width=.9pt},
  atomB/.style={draw=figInk,fill=white,line width=1.7pt},
  bond/.style={draw=figBond,line width=1.9pt}]
\path[use as bounding box] (-0.15,-1.55) rectangle (9.95,4.35);

\node[anchor=west,font=\sffamily\bfseries\fontsize{10}{12}\selectfont] at (0,4.00) {(a) Primitive cell};
\node[anchor=west,font=\sffamily\bfseries\fontsize{10}{12}\selectfont] at (5.35,4.00) {(b) Bloch phases};
\draw[bond] (1.0500,1.2572) -- (1.4500,0.5644);
\draw[bond] (1.0500,1.2572) -- (1.4500,1.9500);
\draw[bond] (1.0500,2.6428) -- (1.4500,1.9500);
\draw[bond] (1.0500,2.6428) -- (1.4500,3.3356);
\draw[bond] (2.2500,0.5644) -- (2.6500,1.2572);
\draw[bond] (2.2500,0.5644) -- (1.4500,0.5644);
\draw[bond] (2.2500,1.9500) -- (2.6500,1.2572);
\draw[bond] (2.2500,1.9500) -- (2.6500,2.6428);
\draw[bond] (2.2500,1.9500) -- (1.4500,1.9500);
\draw[bond] (2.2500,3.3356) -- (2.6500,2.6428);
\draw[bond] (2.2500,3.3356) -- (1.4500,3.3356);
\draw[bond] (3.4500,1.2572) -- (3.8500,0.5644);
\draw[bond] (3.4500,1.2572) -- (3.8500,1.9500);
\draw[bond] (3.4500,1.2572) -- (2.6500,1.2572);
\draw[bond] (3.4500,2.6428) -- (3.8500,1.9500);
\draw[bond] (3.4500,2.6428) -- (3.8500,3.3356);
\draw[bond] (3.4500,2.6428) -- (2.6500,2.6428);
\draw[bond] (4.6500,0.5644) -- (3.8500,0.5644);
\draw[bond] (4.6500,1.9500) -- (3.8500,1.9500);
\draw[bond] (4.6500,3.3356) -- (3.8500,3.3356);
\draw[phaseBlue,line width=1.8pt,dash pattern=on 4pt off 3pt] (1.8500,1.2572) -- (3.0500,0.5644) -- (3.0500,1.9500) -- (1.8500,2.6428) -- cycle;
\path[atomA] (1.0500,1.2572) circle[radius=2.9pt];
\path[atomA] (1.0500,2.6428) circle[radius=2.9pt];
\path[atomA] (2.2500,0.5644) circle[radius=2.9pt];
\path[atomA] (2.2500,1.9500) circle[radius=2.9pt];
\path[atomA] (2.2500,3.3356) circle[radius=2.9pt];
\path[atomA] (3.4500,1.2572) circle[radius=2.9pt];
\path[atomA] (3.4500,2.6428) circle[radius=2.9pt];
\path[atomA] (4.6500,0.5644) circle[radius=2.9pt];
\path[atomA] (4.6500,1.9500) circle[radius=2.9pt];
\path[atomA] (4.6500,3.3356) circle[radius=2.9pt];
\path[atomB] (1.4500,0.5644) circle[radius=2.9pt];
\path[atomB] (1.4500,1.9500) circle[radius=2.9pt];
\path[atomB] (1.4500,3.3356) circle[radius=2.9pt];
\path[atomB] (2.6500,1.2572) circle[radius=2.9pt];
\path[atomB] (2.6500,2.6428) circle[radius=2.9pt];
\path[atomB] (3.8500,0.5644) circle[radius=2.9pt];
\path[atomB] (3.8500,1.9500) circle[radius=2.9pt];
\path[atomB] (3.8500,3.3356) circle[radius=2.9pt];
\draw[white,line width=6pt] (2.2500,1.9500) -- (3.4500,1.2572);
\draw[white,line width=6pt] (2.2500,1.9500) -- (2.2500,3.3356);
\path[atomA] (3.4500,1.2572) circle[radius=2.9pt]; \path[atomA] (2.2500,3.3356) circle[radius=2.9pt];
\draw[vecRed,line width=2.3pt,-{Stealth[length=3.1mm,width=2.5mm]},shorten >=4.2pt] (2.2500,1.9500) -- (3.4500,1.2572);
\draw[vecRed,line width=2.3pt,-{Stealth[length=3.1mm,width=2.5mm]},shorten >=4.2pt] (2.2500,1.9500) -- (2.2500,3.3356);
\path[atomA] (2.2500,1.9500) circle[radius=2.9pt];
\node[fill=white,inner sep=1.2pt,text=vecRed,font=\fontsize{11}{12}\selectfont] at (3.030,1.830) {$\bm a_1$};
\node[fill=white,inner sep=1.2pt,text=vecRed,font=\fontsize{11}{12}\selectfont] at (1.990,2.730) {$\bm a_2$};
\node[fill=white,inner sep=1pt,font=\fontsize{10}{11}\selectfont] at (2.050,1.700) {$A$};
\node[fill=white,inner sep=1pt,font=\fontsize{10}{11}\selectfont] at (2.410,0.987) {$B$};
\draw[figInk,line width=1.4pt,-{Stealth[length=2.2mm]}] (0.30,-0.12) -- (0.95,-0.12) node[right,inner sep=2pt] {$x$};
\draw[figInk,line width=1.4pt,-{Stealth[length=2.2mm]}] (0.30,-0.12) -- (0.30,0.45) node[above,inner sep=2pt] {$y$};
\draw[bond] (7.55,1.95) -- (6.1000,1.9500);
\draw[bond] (7.55,1.95) -- (8.2750,3.2057);
\draw[bond] (7.55,1.95) -- (8.2750,0.6943);
\path[atomB] (6.1000,1.9500) circle[radius=4pt];
\path[atomB] (8.2750,3.2057) circle[radius=4pt];
\path[atomB] (8.2750,0.6943) circle[radius=4pt];
\path[atomA] (7.55,1.95) circle[radius=4.6pt];
\node[font=\fontsize{10}{11}\selectfont] at (7.35,2.37) {$A_{m,n}$};
\node[font=\fontsize{10}{11}\selectfont,anchor=south] at (6.050,2.120) {$B_{m-1,n}$};
\node[font=\fontsize{10}{11}\selectfont,anchor=west] at (8.425,3.286) {$B_{m,n+1}$};
\node[font=\fontsize{10}{11}\selectfont,anchor=west] at (8.425,0.594) {$B_{m,n}$};
\node[text=phaseBlue,fill=white,inner sep=1pt,font=\fontsize{11}{12}\selectfont] at (6.825,1.670) {$e^{-\ii k_X}$};
\node[text=phaseBlue,fill=white,inner sep=1pt,font=\fontsize{11}{12}\selectfont] at (8.463,2.458) {$e^{\ii k_Y}$};
\node[text=phaseBlue,fill=white,inner sep=1pt,font=\fontsize{11}{12}\selectfont] at (8.192,1.302) {$1$};
\node[font=\fontsize{11}{13}\selectfont] at (4.95,-0.85) {$f=-\bigl(1+e^{\ii k_Y}+e^{-\ii k_X}\bigr)$};
\node[font=\fontsize{9}{11}\selectfont,text=figBond] at (4.95,-1.35) {Every microscopic bond: $-t_0$};
\end{tikzpicture}
}

\caption{Real-space origin of the lattice convention.
(a) Filled and open circles denote A and B sites; the dashed primitive
cell contains one of each. Arrows show $\bm a_1$ and $\bm a_2$.
(b) The neighbours of $A_{m,n}$ are $B_{m,n}$, $B_{m,n+1}$ and
$B_{m-1,n}$. Their relative cell-gauge Bloch factors are
$1$, $e^{\ii k_Y}$ and $e^{-\ii k_X}$. Every microscopic hopping
is $-t_0$; the phase factors are not additional hopping strengths.}
\label{fig:latticephases}
\end{figure}
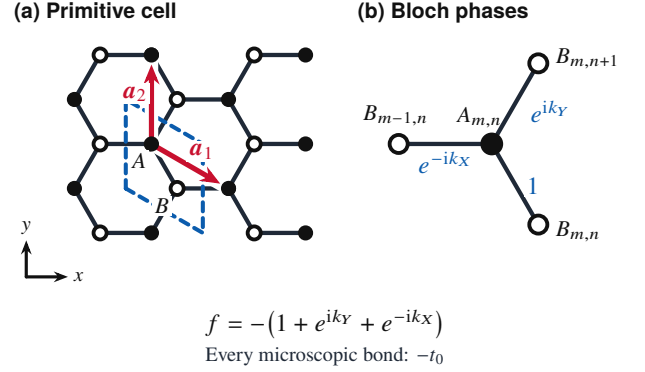

Every angular comparison uses this Cartesian group-velocity angle.
In a warped band it differs from the polar angle of the reduced wave
vector used for the P--N transmission plots in Ref.~\cite{ZhangYang2018}. Angular plots refer to the incident valley
$\xi=+1$; the conductance includes both valleys.
Figure~\ref{fig:latticephases} fixes the cell gauge: the B-site offset
is $a_{\rm CC}(1/2,-\sqrt3/2)$, and both basis sites in cell $(m,n)$
carry the Bloch factor $e^{\ii(mk_X+nk_Y)}$.

\begin{figure*}[!t]
\centering
\definecolor{figInk}{HTML}{111111}
\definecolor{figBond}{HTML}{3B4452}
\definecolor{figNavy}{HTML}{0B2E5C}
\definecolor{figRust}{HTML}{B03A1E}
\definecolor{figTeal}{HTML}{0B7285}
\definecolor{figGrayT}{HTML}{4B5563}
\definecolor{figGrayD}{HTML}{6B7280}
\definecolor{figGrayN}{HTML}{8A9099}
\resizebox{\textwidth}{!}{%
\begin{tikzpicture}[x=1cm,y=1cm,font=\sffamily\fontsize{9}{11}\selectfont,text=figInk,line cap=round,line join=round,
  siteA/.style={draw=figNavy,fill=figNavy,line width=.9pt},
  siteB/.style={draw=figRust,fill=white,line width=1.7pt},
  bond/.style={draw=figBond,line width=1.9pt},
  ptitle/.style={anchor=west,font=\sffamily\bfseries\fontsize{10.5}{12}\selectfont},
  sub/.style={font=\fontsize{8.5}{10}\selectfont,text=figGrayT},
  lab/.style={font=\fontsize{9.5}{11}\selectfont,inner sep=1.5pt}]
\path[use as bounding box] (-0.15,-0.75) rectangle (18.0,9.05);

\node[ptitle] at (0,8.45) {(a) Interface cells};
\node[lab,text=figGrayT] at (2.95,7.85) {$N$: $m\le0$};
\node[lab,text=figGrayT] at (4.85,7.85) {$S$: $m\ge1$};
\draw[figTeal,line width=1.6pt] (2.15,5.50) rectangle (5.70,7.45);
\draw[figGrayD,line width=1.2pt,dash pattern=on 3pt off 2.5pt] (3.925,5.40) -- (3.925,7.55);
\draw[bond] (0.6,6.9) -- (1.2999999999999998,6.3);
\draw[bond] (1.2999999999999998,6.3) -- (2.6,6.9);
\draw[bond] (2.6,6.9) -- (3.3,6.3);
\draw[draw=figRust,line width=2.8pt] (3.3,6.3) -- (4.6,6.9);
\draw[bond] (4.6,6.9) -- (5.3,6.3);
\draw[bond] (5.3,6.3) -- (6.6,6.9);
\draw[bond] (6.6,6.9) -- (7.3,6.3);
\path[siteA] (0.6,6.9) circle[radius=3.4pt];
\path[siteA] (2.6,6.9) circle[radius=3.4pt];
\path[siteA] (4.6,6.9) circle[radius=3.4pt];
\path[siteA] (6.6,6.9) circle[radius=3.4pt];
\path[siteB] (1.2999999999999998,6.3) circle[radius=3.4pt];
\path[siteB] (3.3,6.3) circle[radius=3.4pt];
\path[siteB] (5.3,6.3) circle[radius=3.4pt];
\path[siteB] (7.3,6.3) circle[radius=3.4pt];
\node[lab] at (0.6,7.32) {$A$};
\node[lab] at (1.2999999999999998,5.88) {$B$};
\node[lab] at (0.95,5.15) {$m=-1$};
\node[lab] at (2.6,7.32) {$A$};
\node[lab] at (3.3,5.88) {$B$};
\node[lab] at (2.95,5.15) {$m=0$};
\node[lab] at (4.6,7.32) {$A$};
\node[lab] at (5.3,5.88) {$B$};
\node[lab] at (4.949999999999999,5.15) {$m=1$};
\node[lab] at (6.6,7.32) {$A$};
\node[lab] at (7.3,5.88) {$B$};
\node[lab] at (6.949999999999999,5.15) {$m=2$};
\node[lab,text=figRust] at (3.925,4.45) {Interface bond: $B_0$--$A_1$};
\node[ptitle] at (9.4,8.45) {(b) A finite normal mode};
\node[sub] at (13.600000000000001,7.95) {$j=\mathrm{in},e^-,h^-$;\quad $c_j=1,r_e,r_h$};
\draw[bond] (10.3,6.9) -- (11.850000000000001,6.3);
\draw[bond] (11.850000000000001,6.3) -- (14.9,6.9);
\draw[bond] (14.9,6.9) -- (16.4,6.3);
\path[siteA] (10.3,6.9) circle[radius=3.4pt];
\path[siteA] (14.9,6.9) circle[radius=3.4pt];
\path[siteB] (11.850000000000001,6.3) circle[radius=3.4pt];
\path[siteB] (16.4,6.3) circle[radius=3.4pt];
\node[lab,text=figNavy] at (10.15,7.4) {$c_j\lambda_j^{-1}\bm q_{A,j}$};
\node[lab,text=figRust] at (11.750000000000002,5.8) {$c_j\lambda_j^{-1}\bm q_{B,j}$};
\node[lab,text=figNavy] at (14.9,7.4) {$c_j\bm q_{A,j}$};
\node[lab,text=figRust] at (16.4,5.8) {$c_j\bm q_{B,j}$};
\node[lab] at (11.075000000000001,5.15) {$m=-1$};
\node[lab] at (15.649999999999999,5.15) {$m=0$};
\node[font=\fontsize{10.5}{12}\selectfont] at (13.600000000000001,4.45) {$\Psi^{N,j}_m=c_j\lambda_j^{m}\Phi_j,\quad m\le0$};
\node[ptitle] at (0.0,3.8999999999999995) {(c) A finite superconducting mode};
\node[sub] at (4.2,3.4000000000000004) {$n=1,2$;\quad $|\lambda_{Sn}|<1$};
\draw[bond] (0.9,2.3500000000000005) -- (2.45,1.75);
\draw[bond] (2.45,1.75) -- (5.5,2.3500000000000005);
\draw[bond] (5.5,2.3500000000000005) -- (7.0,1.75);
\path[siteA] (0.9,2.3500000000000005) circle[radius=3.4pt];
\path[siteA] (5.5,2.3500000000000005) circle[radius=3.4pt];
\path[siteB] (2.45,1.75) circle[radius=3.4pt];
\path[siteB] (7.0,1.75) circle[radius=3.4pt];
\node[lab,text=figNavy] at (0.8,2.8500000000000005) {$t_n\bm q_{A,n}$};
\node[lab,text=figRust] at (2.45,1.25) {$t_n\bm q_{B,n}$};
\node[lab,text=figNavy] at (5.5,2.8500000000000005) {$t_n\lambda_{Sn}\bm q_{A,n}$};
\node[lab,text=figRust] at (7.0,1.25) {$t_n\lambda_{Sn}\bm q_{B,n}$};
\node[lab] at (1.675,0.6000000000000005) {$m=1$};
\node[lab] at (6.25,0.6000000000000005) {$m=2$};
\node[font=\fontsize{10.5}{12}\selectfont] at (4.2,-0.09999999999999964) {$\Psi^{S,n}_m=t_n\lambda_{Sn}^{m-1}\Phi_{Sn},\quad m\ge1$};
\node[ptitle] at (9.4,3.8999999999999995) {(d) Candidate boundary amplitudes};
\node[sub] at (13.600000000000001,3.4000000000000004) {Additional contribution only};
\draw[draw=figGrayN,line width=1.9pt] (10.3,2.3500000000000005) -- (11.850000000000001,1.75);
\draw[bond] (11.850000000000001,1.75) -- (14.9,2.3500000000000005);
\draw[draw=figGrayN,line width=1.9pt] (14.9,2.3500000000000005) -- (16.4,1.75);
\draw[figGrayD,line width=1.2pt,dash pattern=on 3pt off 2.5pt] (13.375,1.0) -- (13.375,2.9000000000000004);
\path[draw=figGrayN,fill=figGrayN,line width=.9pt] (10.3,2.3500000000000005) circle[radius=3.4pt];
\path[siteB] (11.850000000000001,1.75) circle[radius=3.4pt];
\path[draw=figRust,fill=figRust,line width=.9pt] (14.9,2.3500000000000005) circle[radius=3.4pt];
\path[draw=figGrayN,fill=white,line width=1.7pt] (16.4,1.75) circle[radius=3.4pt];
\node[lab,text=figGrayT] at (10.3,2.8000000000000007) {$0$};
\node[lab,text=figRust] at (11.700000000000001,1.2) {$(a_{L,e},a_{L,h})^{\mathsf T}$};
\node[lab,text=figRust] at (15.25,2.8500000000000005) {$(a_{R,e},a_{R,h})^{\mathsf T}$};
\node[lab,text=figGrayT] at (16.4,1.3) {$0$};
\node[lab] at (11.075000000000001,0.6000000000000005) {$m=0$};
\node[lab] at (15.649999999999999,0.6000000000000005) {$m=1$};
\node[lab,text=figRust] at (13.600000000000001,-0.09999999999999964) {All four vanish for unchanged hopping ($\eta\ne0$).};
\end{tikzpicture}
}
\caption{Site amplitudes in the lattice matching construction.
Write $\bm q_{s,j}=(u_{s,j},v_{s,j})^{\mathsf T}$ for the electron
and hole components on sublattice $s=A,B$ of mode $j$.
(a) The reduced chain is drawn schematically. The outline encloses cells
0 and 1, and the highlighted bond joins $B_0$ to $A_1$.
(b) One finite normal mode, referenced to cell 0; $j=\mathrm{in},e^-,h^-$
has coefficient $c_j=1,r_e,r_h$, respectively.
(c) One right-decaying S mode, referenced to cell 1.
(d) Candidate additional boundary contributions have support on $B_0$
and $A_1$ only. Zeros refer to this additional contribution, not to the
full state. All four coefficients vanish for the present unchanged-hopping
step with $\eta\ne0$, as shown in Eq.~\eqref{eq:vanishing}.}
\label{fig:matchingstates}
\end{figure*}
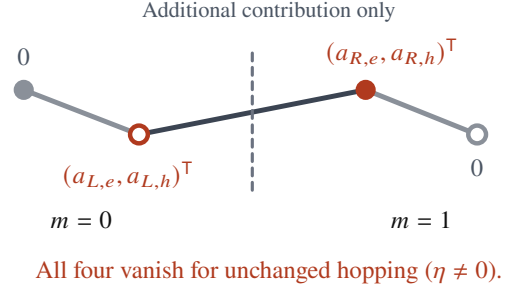

\subsection{Lattice modes at fixed transverse momentum}

We use the lead-mode formulation of lattice transport
\cite{Ando1991,Khomyakov2005,Sorensen2009}. Its connection to
Green-function scattering theory is established by the
conductivity--transmission relation and extensions to layered,
multiorbital systems
\cite{FisherLee1981,Sanvito1999,ZhangYangChang2017}.
Fourier transformation along the interface leaves a chain labelled by
the cell index $m$ [Fig.~\ref{fig:model}(b)]. Its electronic on-cell
and intercell matrices, in units of $t_0$, are
\begin{equation}
 h=\begin{pmatrix}0&-\eta\\-\eta^*&0\end{pmatrix},\quad
 t=\begin{pmatrix}0&0\\-1&0\end{pmatrix},\quad \eta=1+e^{\ii k_Y}.
\end{equation}
The two terms in $\eta$ account for the bonds within the effective
cell, while $t$ connects $B_m$ to $A_{m+1}$. Its zero entries
follow from the cell convention, not from missing graphene bonds.
The normal-state cell equation is
\begin{equation}
 t\psi_{m+1}+h\psi_m+t^\dagger\psi_{m-1}=\epsilon\psi_m.
\end{equation}
For $\lambda=e^{\ii k_X}$, the Bloch Hamiltonian is
$H_0=t\lambda+h+t^\dagger\lambda^{-1}$. The off-diagonal structure factor is
$f(\bm k)=-(1+e^{\ii k_Y}+e^{-\ii k_X})$, giving
$\epsilon_\pm=\pm|f|$ in the uniform normal lattice.
At a chosen normal-state energy $\epsilon$, the finite longitudinal roots satisfy
\begin{equation}
 \eta\lambda^2+(3+2\cos k_Y-\epsilon^2)\lambda+\eta^*=0.
 \label{eq:bloch}
\end{equation}
For complex Bloch factors, define the matrix entries separately:
\begin{equation}
 f=-(\eta+\lambda^{-1}),\qquad g=-(\eta^*+\lambda).
\end{equation}
The secular equation $fg=\epsilon^2$ gives
$|\eta|^2+1+\eta\lambda+\eta^*/\lambda=\epsilon^2$.
Multiplication by $\lambda$ gives Eq.~\eqref{eq:bloch}, with roots
\begin{equation}
 \lambda_\pm=\frac{-C\pm\sqrt{C^2-4|\eta|^2}}{2\eta},\quad
 C=3+2\cos k_Y-\epsilon^2.
 \label{eq:rootsTB}
\end{equation}
The exceptional point $\eta=0$ is outside the normal-lead valley
intervals used here. A normalized sublattice spinor is
\begin{equation}
 u(\epsilon,\lambda)=
 \frac{(f,\epsilon)^{\mathsf T}}{\sqrt{|f|^2+|\epsilon|^2}}.
 \label{eq:spinor}
\end{equation}
The equivalent chart $(\epsilon,g)^{\mathsf T}$ is chosen if it
has the larger norm. This avoids losing an eigenvector because
of a poor choice of components. For a propagating mode $g=f^*$;
for an evanescent mode $g$ must retain its analytic expression.
Numerically conjugating $f$ at complex $\lambda$ would solve a
different eigenvalue problem.

We use the time-reversed Nambu basis
\begin{equation}
 \Psi=(\psi_{Ae},\psi_{Be},\psi_{Ah},\psi_{Bh})^{\mathsf T}.
\end{equation}
The hole represents a missing opposite-spin electron in the opposite valley.
Thus the singlet pairing is intervalley even though the same sublattice
matrix appears in both diagonal BdG blocks
\cite{Beenakker2006,AkhmerovBeenakker2007}. The BdG cell and hopping matrices are
\begin{equation}
 H_j=\begin{pmatrix}h-\mu_j\Id_2&\Delta_j\Id_2\\
 \Delta_j^*\Id_2&\mu_j\Id_2-h\end{pmatrix},\qquad
 \mathcal T=\begin{pmatrix}t&0\\0&-t\end{pmatrix}.
\end{equation}
The electron and hole roots in N follow by setting
$\epsilon=\mu_N+E$ and $\mu_N-E$, respectively, in
Eq.~\eqref{eq:bloch}. In S the corresponding targets are
$\epsilon=\mu_S\pm\ii\Omega$, where
$\Omega=\sqrt{\Delta_0^2-E^2}$.
The normal spinors are $\Phi_e=(u_e,0_2)^{\mathsf T}$ and
$\Phi_h=(0_2,u_h)^{\mathsf T}$. The hole current has the opposite
sign to the electronic band velocity of $u_h$.

To construct the S modes, first take $H_0u=\epsilon u$. The
remaining electron--hole eigenproblem is
\begin{equation}
 \begin{pmatrix}\epsilon-\mu_S&\Delta_S\\
 \Delta_S^*&\mu_S-\epsilon\end{pmatrix}
 \begin{pmatrix}c_e\\c_h\end{pmatrix}
 =E\begin{pmatrix}c_e\\c_h\end{pmatrix}.
\end{equation}
Its determinant gives $(\epsilon-\mu_S)^2=E^2-|\Delta_S|^2$.
With $\epsilon_\sigma=\mu_S+\sigma\ii\Omega$, the spinor is
\begin{equation}
 \Phi_{S\sigma}=\frac{(u_\sigma,\gamma_\sigma u_\sigma)^{\mathsf T}}
 {\sqrt{1+|\gamma_\sigma|^2}},\qquad
 \gamma_\sigma=\frac{E-\sigma\ii\Omega}{\Delta_S}.
 \label{eq:Sspinor}
\end{equation}
For each $\sigma=\pm1$, one lattice root decays into S.
The two such roots form the superconducting basis. Their
Euclidean normalization is a convention, not a unit-flux
normalization of a decaying state.

Unit-modulus roots propagate and their direction is determined
by the BdG current. A root with $|\lambda|<1$ decays to the
right; one with $|\lambda|>1$ decays to the left. When the
reflected hole channel closes, its left-decaying continuation
remains in the matching basis. Its presence affects the boundary
solution even though its asymptotic reflected flux is zero.

Figure~\ref{fig:matchingstates} shows the finite-mode amplitudes,
referenced to cell 0 in N and cell 1 in S. Panel (d) isolates candidate
boundary contributions; their coefficients follow from the interface
recurrences.

\subsection{Interface matching and reflection probabilities}
Because $\mathcal T$ is singular, its inverse cannot be used to define
an ordinary transfer matrix. Singular or ill-conditioned intercell
couplings are a known issue in lead calculations
\cite{RunggerSanvito2008}. Here we impose the two interface-cell
recurrences,
\begin{align}
 (H_N-E)\Psi_0+\mathcal T^\dagger\Psi_{-1}+\mathcal T\Psi_1&=0,\notag\\
 \mathcal T^\dagger\Psi_0+(H_S-E)\Psi_1+\mathcal T\Psi_2&=0.
 \label{eq:boundary}
\end{align}
Let $m=0$ be the last normal cell and $m=1$ the first
superconducting cell. The finite-mode scattering state is
\begin{align}
 \Psi_m^N&=\lambda_i^m\Phi_i+
 r_e\lambda_e^m\Phi_e^-+r_h\lambda_h^m\Phi_h^-,\quad m\leq0,
 \notag\\
 \Psi_m^S&=\sum_{n=1}^2t_n\lambda_{Sn}^{m-1}\Phi_{Sn},\quad m\geq1.
 \label{eq:states}
\end{align}
The right-hand modes are referenced to cell 1, so their
extrapolated values at cell 0 contain $\lambda_{Sn}^{-1}$.
This convention accounts for the inverse Bloch factors
in the final matching matrix.

An augmented basis initially includes possible amplitudes confined to
the boundary cells. For the unchanged-hopping step and $\eta\ne0$,
four recurrence rows force all these additional amplitudes to vanish.
At cell 0 the allowed vectors belong to $\ker\mathcal T$
and have only $B_e$ or $B_h$ components. At cell 1 they belong
to $\ker\mathcal T^\dagger$ and have only $A_e$ or $A_h$
components. Denote their coefficients by
$a_{L,e},a_{L,h},a_{R,e},a_{R,h}$. The corresponding rows give
\begin{align}
 -\eta a_{L,e}=0,&\qquad \eta a_{L,h}=0,\notag\\
 -\eta^*a_{R,e}=0,&\qquad \eta^*a_{R,h}=0.
 \label{eq:vanishing}
\end{align}
The finite modes cancel from these rows by their bulk equations.
Since $\eta\ne0$, all four extra coefficients are zero.
The surviving conditions match the extrapolated $A$ components
at cell 1 and $B$ components at cell 0. They match the lattice
state across the interface bond.
The remaining system is
\begin{equation}
 M_4(r_e,r_h,t_1,t_2)^{\mathsf T}
 =-(\lambda_i A_i,0,B_i,0)^{\mathsf T},
\end{equation}
with
\begin{equation}
 M_4=\begin{pmatrix}
 \lambda_e A_e&0&-A_1&-A_2\\
 0&\lambda_h C_h&-C_1&-C_2\\
 B_e&0&-B_1/\lambda_{S1}&-B_2/\lambda_{S2}\\
 0&D_h&-D_1/\lambda_{S1}&-D_2/\lambda_{S2}
 \end{pmatrix}.
 \label{eq:m4}
\end{equation}
Here $(A_i,B_i)$ is the incident electron spinor,
$(A_e,B_e)$ and $(C_h,D_h)$ are the reflected electron and hole
spinors, and $(A_n,B_n,C_n,D_n)$ are the two decaying S spinors.
The reduction is specific to this interface. A changed contact
hopping requires the boundary equations to be derived again.
The full augmented construction is retained in the supplement.

\begin{figure}[!t]
\centering\includegraphics[width=\columnwidth]{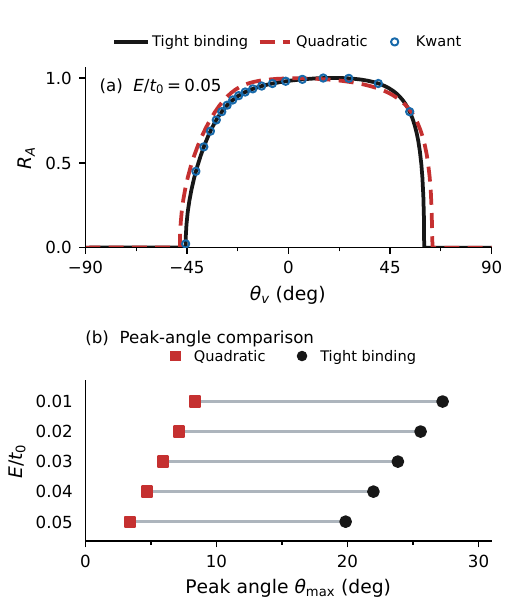}
\caption{Angular consequence of interface matching for
$(\mu_N,\mu_S,\Delta_0)/t_0=(0.5,1,0.1)$, $\xi=+1$.
(a) Andreev reflection versus the group-velocity incidence angle
at $E/t_0=0.05$: tight binding (black solid), quadratic continuum
(red dashed), and selected independent Kwant values (open blue circles).
(b) Refined peak angles at five excitation energies. Black circles
and red squares denote the tight-binding and quadratic results,
respectively. Each grey segment joins the two predictions at the
same energy. The five full angular sweeps appear in Fig.~S1.}
\label{fig:finite}
\end{figure}

\subsection{Reflection amplitudes and the superconducting boundary}
The four-equation system also gives an analytical view of
the conversion process. Define
\begin{equation}
 U_A=\begin{pmatrix}A_1&A_2\\C_1&C_2\end{pmatrix},\quad
 U_B=\begin{pmatrix}
 B_1/\lambda_{S1}&B_2/\lambda_{S2}\\
 D_1/\lambda_{S1}&D_2/\lambda_{S2}
 \end{pmatrix}.
\end{equation}
If $U_A$ is invertible, let $Y=U_BU_A^{-1}$. Eliminating the
two superconducting coefficients gives
\begin{equation}
 \begin{pmatrix}B_i+r_eB_e\\r_hD_h\end{pmatrix}
 =Y\begin{pmatrix}\lambda_iA_i+r_e\lambda_eA_e\\
 r_h\lambda_hC_h\end{pmatrix}.
 \label{eq:Ymap}
\end{equation}
With $L_e=B_e-Y_{ee}\lambda_eA_e$ and
$L_h=D_h-Y_{hh}\lambda_hC_h$, the reflection amplitudes are
\begin{align}
 \mathcal D&=L_eL_h-Y_{eh}Y_{he}\lambda_e\lambda_hA_eC_h,\notag\\
 r_h&=\frac{Y_{he}(\lambda_iA_iB_e-\lambda_eA_eB_i)}{\mathcal D},
 \label{eq:rh}\\
 r_e&=\frac{(Y_{ee}\lambda_iA_i-B_i)L_h+
 Y_{eh}Y_{he}\lambda_hC_h\lambda_iA_i}{\mathcal D}.
 \label{eq:re}
\end{align}
These expressions require nonsingular $U_A$ and $\mathcal D$.
The production calculation uses a direct solve of
Eq.~\eqref{eq:m4}; an inconvenient elimination is not evidence
of a physical singularity.

The matrix $Y$ contains both decaying S modes and their
sublattice phases. The incident and reflected normal spinors
enter separately. Andreev reflection therefore depends on the
superconducting boundary response as well as the normal-state
contours. A change in the incident dispersion alone cannot fix
the peak shift. Equations~\eqref{eq:rh} and \eqref{eq:re} do
not establish a universal pseudospin-conservation rule for unit
Andreev reflection.

\subsection{Current normalization}
The conserved BdG current is a probability current. Electrical
current also accounts for the opposite charge of a reflected
hole. With dimensionless time $\tau=t_0t_{\rm phys}/\hbar$
and density $\rho_m=\Psi_m^\dagger\Psi_m$, the cell equation gives
\begin{align}
 \partial_\tau\rho_m&=\mathcal J_{m-1\to m}-\mathcal J_{m\to m+1},
 \notag\\
 \mathcal J_{m\to m+1}&=-2\operatorname{Im}
 (\Psi_m^\dagger\mathcal T\Psi_{m+1}).
 \label{eq:bondcurrent}
\end{align}
The Hermitian on-cell terms cancel from this continuity
equation. For a propagating Bloch wave, the bond current becomes
$\Phi^\dagger(\ii\mathcal T\lambda-
\ii\mathcal T^\dagger\lambda^{-1})\Phi$. The matrix in
parentheses is the derivative of the Bloch Hamiltonian.

For a propagating mode $\Phi_\alpha$, therefore,
\begin{equation}
 J_\alpha=\Phi_\alpha^\dagger\frac{\partial H_{\rm BdG}}
 {\partial k_X}\Phi_\alpha,\qquad
 R_{N,A}=|r_{e,h}|^2\frac{|J_{e,h}^{-}|}{J_{\rm in}}.
 \label{eq:flux}
\end{equation}
For continuum modes, the derivative is with respect to $p_X$.
The common dimensional factor cancels in the probability ratios.
Evanescent modes enter the boundary solution but carry no asymptotic
reflected flux. For the open subgap channels, $R_N+R_A=1$.
Rescaling a basis spinor changes its coefficient inversely,
leaving these probabilities unchanged. Thus $|r_h|^2$ alone
is not a reflection probability for unit-norm modes. At a
channel threshold the longitudinal current vanishes, so the
exact endpoint is excluded from flux normalization.
\subsection{Continuum models and interface conditions}
Write $(k_X,k_Y)=(\xi K_0+p_X,\xi K_0+p_Y)$, with $K_0=2\pi/3$.
Expanding the nearest-neighbour structure factor to second order gives
\begin{align}
 f_\xi={}&\ii e^{-\ii\xi K_0}p_X-\ii e^{\ii\xi K_0}p_Y\notag\\
 &+\tfrac12(e^{-\ii\xi K_0}p_X^2+e^{\ii\xi K_0}p_Y^2).
 \label{eq:fquad}
\end{align}
The partner $\bar f_\xi$ is the analytic continuation of $f_\xi^*$
from real momenta. At complex $p_X$, it must not be evaluated by
numerically conjugating $f_\xi$. Set
$h_\xi=\bigl(\begin{smallmatrix}0&f_\xi\\\bar f_\xi&0\end{smallmatrix}\bigr)$
and $F_\xi=f_\xi\bar f_\xi$.

For real momenta, the leading trigonal warping is explicit in
Cartesian coordinates. Define $Q_x=a_{\rm CC}q_x$,
$Q_y=a_{\rm CC}q_y$, and $(Q_x,Q_y)=Q(\cos\phi,\sin\phi)$.
Substituting $p_X=(3Q_x-\sqrt3Q_y)/2$ and $p_Y=\sqrt3Q_y$
into $F_\xi$ gives
\begin{align}
 F_\xi&=\frac94Q^2-\frac{9\xi}{8}Q^3\sin(3\phi)+O(Q^4),
 \label{eq:trigonal}\\
 \epsilon_+&=\frac32Q\left[1-\frac{\xi Q}{4}\sin(3\phi)
 +O(Q^2)\right].
 \label{eq:warpedband}
\end{align}
Here $\epsilon_+$ is measured from the Dirac point in units of $t_0$.
The threefold term has the same leading coefficient in the exact
lattice and quadratic models and reverses sign between valleys.
The sine, rather than a cosine, follows from our lattice orientation.
Because the group velocity is normal to a constant-energy contour,
warping also changes its direction relative to the wave vector.
For an incident conduction-band electron,
\begin{equation}
 \theta_v-\phi=-\frac{3\xi Q}{4}\cos(3\phi)+O(Q^2),
 \label{eq:velocitydeflection}
\end{equation}
with the angular difference in radians. This is why the physical
incidence angle $\theta_v$ and the momentum angle $\phi$ must be
distinguished in a comparison with Ref.~\cite{ZhangYang2018}.

\begin{figure}[!t]
\centering
% Single-column vector figure: exact contour, white background.
\definecolor{warpInk}{HTML}{111111}
\definecolor{warpRed}{HTML}{D20A11}
\definecolor{warpBlue}{HTML}{0057B8}
\resizebox{\columnwidth}{!}{%
\begin{tikzpicture}[font=\sffamily\fontsize{9}{11}\selectfont,
  text=warpInk,line cap=round,line join=round,
  axis/.style={draw=warpInk,line width=.7pt,-{Stealth[length=1.7mm]}},
  momentum/.style={draw=warpBlue,line width=2.1pt,
    -{Stealth[length=2.6mm,width=2.0mm]}},
  velocity/.style={draw=warpInk,line width=2.1pt,
    -{Stealth[length=2.6mm,width=2.0mm]}}]
\path[use as bounding box] (0,0) rectangle (8.4,8.05);
\def\twEnergy{0.8}
\pgfmathsetmacro{\twRadius}{2*\twEnergy/3}
\pgfmathsetmacro{\twLo}{2*acos((1+\twEnergy)/2)}
\pgfmathsetmacro{\twHi}{2*acos((1-\twEnergy)/2)}
\pgfmathsetmacro{\twMid}{(\twLo+\twHi)/2}
\pgfmathsetmacro{\twSpan}{(\twHi-\twLo)/2}
\pgfmathsetmacro{\twPoint}{2/3*rad(acos(1-\twEnergy^2/2))}
\pgfmathsetmacro{\twAngle}{-asin(\twEnergy/2)}
% PGF trigonometric functions use degrees; rad() restores physical momenta.
\tikzset{declare function={
  twK(\u)=\twMid+\twSpan*cos(\u);
  twQ(\u)=2*cos(twK(\u)/2);
  twX(\u)=2/3*rad(acos(min(1,max(-1,
    (1+twQ(\u)^2-\twEnergy^2)/(2*twQ(\u))))));
  twY(\u)=rad(twK(\u)-120)/sqrt(3);
}}

% One shared set of axes makes the threefold distortion easy to compare.
\draw[warpInk,line width=1.4pt,dash pattern=on 4pt off 3pt]
  (.45,7.65) -- (1.20,7.65);
\node[anchor=west,font=\sffamily\fontsize{8.5}{10}\selectfont]
  at (1.34,7.65) {Linear Dirac};
\draw[warpRed,line width=2.2pt] (4.25,7.65) -- (5.00,7.65);
\node[anchor=west,font=\sffamily\fontsize{8.5}{10}\selectfont]
  at (5.14,7.65) {Exact lattice};
\node[font=\fontsize{9}{11}\selectfont] at (4.05,7.10)
  {$\xi=+1,\qquad\epsilon_+=0.8$};
\begin{scope}[shift={(3.25,3.65)},x=4cm,y=4cm]
  \draw[axis] (-.76,0) -- (1.03,0) node[right=2pt] {$Q_x$};
  \draw[axis] (0,-.82) -- (0,.74) node[right=3pt] {$Q_y$};
  \draw[warpInk,line width=1.4pt,dash pattern=on 4pt off 3pt]
    (0,0) circle[radius=\twRadius];
  \draw[warpRed,line width=2.2pt]
    plot[domain=0:180,samples=181,variable=\u] ({twX(\u)},{twY(\u)})
    -- plot[domain=180:0,samples=181,variable=\u] ({-twX(\u)},{twY(\u)}) -- cycle;
  \coordinate (P) at (\twPoint,0);
  % Exact tangent and outward velocity at Q_y=0, Q_x>0.
  \draw[warpInk,line width=1pt]
    (P) ++({.18*cos(\twAngle+90)},{.18*sin(\twAngle+90)})
    -- ++({-.36*cos(\twAngle+90)},{-.36*sin(\twAngle+90)});
  \draw[momentum] (0,0) -- (P)
    node[midway,above=4pt,text=warpBlue] {$\bm q$};
  \draw[velocity] (P) -- ++({.34*cos(\twAngle)},{.34*sin(\twAngle)})
    node[right=3pt] {$\bm v$};
  \fill[warpInk] (0,0) circle[radius=.015];
  \fill[warpInk] (P) circle[radius=.018];
  \node[below left=2pt] at (0,0) {$0$};
  \node at (.67,-.59) {$\bm v\nparallel\bm q$};
\end{scope}
\end{tikzpicture}%
}
\caption{Trigonal warping and the direction of propagation.
The red solid curve is the exact nearest-neighbour contour in
valley $\xi=+1$ at normal-state energy $\epsilon_+=0.8$ in units
of $t_0$; the black dashed circle is its linear Dirac reference.
Here $(Q_x,Q_y)=a_{\rm CC}(q_x,q_y)$ is measured from the valley
centre. At $Q_y=0$, $Q_x>0$, the blue arrow shows $\bm q$ and
the black arrow shows the velocity direction. Velocity is normal
to the contour (the short black segment is tangent), so it is
not parallel to $\bm q$. Arrow lengths are illustrative.
The energy is chosen to display the geometry, not to specify
another NS transport calculation.}
\label{fig:trigonalgeometry}
\end{figure}
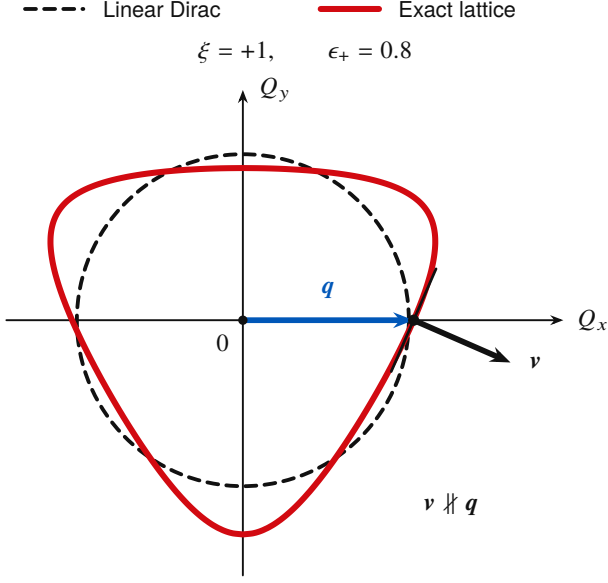

Figure~\ref{fig:trigonalgeometry} illustrates this geometric
distinction using the exact lattice contour.

The regional BdG Hamiltonian is
\begin{equation}
 \mathcal H_{j\xi}=\begin{pmatrix}
 h_\xi-\mu_j\Id_2&\Delta_j\Id_2\\
 \Delta_j^*\Id_2&\mu_j\Id_2-h_\xi
 \end{pmatrix}.
\end{equation}
Since $h_\xi^2=F_\xi\Id_2$, the electronic eigenvalues are
$\epsilon=\pm\sqrt{F_\xi}$. In each electronic eigenspace,
$E^2=(\epsilon-\mu_j)^2+|\Delta_j|^2$. Multiplying the two
factors for the signs of $\epsilon$ removes the square root.
The longitudinal roots then follow from
\begin{equation}
 (F_\xi+\mu_j^2+|\Delta_j|^2-E^2)^2-4\mu_j^2F_\xi=0.
 \label{eq:det}
\end{equation}
Equivalently, solve $F_\xi=\Lambda^2$ with
$\Lambda=\mu_N\pm E$ in N and
$\Lambda=\mu_S\pm\ii\Omega$ in S, where
$\Omega=\sqrt{\Delta_0^2-E^2}$. Each equation is a quartic in $p_X$.
Current and decay direction select the outgoing and decaying modes.

Here $X$ is the dimensionless coordinate conjugate to $p_X$.
For fixed $p_Y$, the differential operator has the form
\begin{equation}
 \mathcal H=A_2(-\ii\partial_X)^2+A_1(-\ii\partial_X)+A_0(X).
\end{equation}
Only $A_0$ contains the finite steps. The highest derivative
coefficient is
\begin{equation}
 A_2=\operatorname{diag}(h_2,-h_2),\qquad
 h_2=\frac12\begin{pmatrix}0&e^{-\ii\xi K_0}\\
 e^{\ii\xi K_0}&0\end{pmatrix}.
\end{equation}
It obeys $A_2^2=\Id_4/4$ and is invertible. Both derivative
coefficients are the same on the two sides.

Allow a possible wave-function jump
$[\Psi]=\Psi(0^+)-\Psi(0^-)$. Its first derivative contains
$[\Psi]\delta(X)$; its second derivative contains
$[\Psi']\delta(X)+[\Psi]\delta'(X)$. The coefficients of
these singular terms must vanish:
\begin{equation}
 -A_2[\Psi]=0,\qquad -A_2[\Psi']-\ii A_1[\Psi]=0.
\end{equation}
There is no singular potential to cancel them. Invertibility
of $A_2$ gives first $[\Psi]=0$ and then $[\Psi']=0$.
The interface conditions are therefore
\begin{equation}
 \Psi_N(0)=\Psi_S(0),\qquad
 \partial_X\Psi_N(0)=\partial_X\Psi_S(0).
 \label{eq:qboundary}
\end{equation}
Four normal outgoing or left-decaying modes and four right-decaying
superconducting modes give eight unknown coefficients. They are found
from these eight scalar conditions. A mode referenced to $X=0$ contributes the boundary column
$(\chi,\ii p_X\chi)^{\mathsf T}$, with opposite signs for
normal and superconducting columns. The negative incident
column is the source. The polynomials and complete matrix are
given in the Supplemental Material.

The continuum probability current is
\begin{equation}
 J_{\rm QC}=\Psi^\dagger A_1\Psi+
 2\operatorname{Im}(\Psi^\dagger A_2\partial_X\Psi).
\end{equation}
For real $p_X$, it becomes
$\chi^\dagger(A_1+2p_XA_2)\chi$. The matching conditions
make this current continuous, verifying the consistency of
the specified differential boundary problem.

Equation~\eqref{eq:qboundary} specifies a continuum interface; it is not
a microscopic derivation of the lattice boundary. For example, at
$p_Y=0$,
\begin{equation}
 F_\xi=p_X^2+\tfrac14p_X^4,\qquad
 p_X^2=-2\pm2\sqrt{1+\Lambda^2}.
 \label{eq:auxiliary}
\end{equation}
For real $\Lambda\geq0$, the positive branches are
\begin{align}
 p_{\rm phys}&=\sqrt{-2+2\sqrt{1+\Lambda^2}},\notag\\
 \kappa_{\rm aux}&=\sqrt{2+2\sqrt{1+\Lambda^2}},\qquad
 p_{\rm aux}=\ii\kappa_{\rm aux}.
 \label{eq:branches}
\end{align}
The negative roots complete both pairs. For small $\Lambda$,
$p_{\rm phys}=\Lambda-\Lambda^3/8+O(\Lambda^5)$ and
$\kappa_{\rm aux}=2+\Lambda^2/4+O(\Lambda^4)$.
The physical roots therefore approach the valley, while the auxiliary
roots approach $\pm2\ii$ and retain a decay length of order a lattice
spacing. These short-wavelength modes are required by the second-order
boundary problem but lie outside the controlled valley expansion.
This boundary limitation applies to all continuum--lattice comparisons
below.

Dropping the quadratic terms gives the linear-Dirac model.
Its first-order equation requires continuity of $\Psi$ alone;
adding derivative continuity would overconstrain that problem.
The relation between a microscopic graphene boundary and its
continuum condition is also important for terminated ribbons
\cite{BreyFertig2006,AkhmerovBeenakker2008Boundary}, although
a terminated edge differs from the transmitting NS interface.
Here each boundary problem is stated explicitly before its
predictions are compared.

\section{Results and discussion}

\subsection{Angular reflection at finite doping}
Figure~\ref{fig:finite} collects the finite-doping comparison. At
$(\mu_N,\mu_S,\Delta_0)/t_0=(0.5,1,0.1)$, both models show broad
regions of high Andreev reflection. Their maxima differ: at
$E/t_0=0.05$, the peak is at $19.86^\circ$ in the lattice model and
$3.43^\circ$ in the quadratic model. As $E/t_0$ increases from 0.01
to 0.05, both peaks move toward normal incidence.
Figure~\ref{fig:finite}(b) pairs the two peak angles at each energy.
Their separation decreases from $18.88^\circ$ to $16.43^\circ$
over this range, so a substantial angular difference remains.
The full angular sweeps are retained in Fig.~S1.
The broad high-reflection region does not imply unit reflection: the
refined maxima at this doping remain below one.
At $E/t_0=0.01$, for example, they are 0.99692 and 0.99467.
A broad maximum is difficult to locate by eye even when the
two peak angles differ substantially; its position and height
are therefore reported separately.

Trigonal warping provides a physical source of angular asymmetry
within a single valley. A fixed $k_Y$ cuts the warped electron
and hole contours at different longitudinal momenta, selecting
different sublattice spinors and current magnitudes. Equations
\eqref{eq:trigonal} and \eqref{eq:velocitydeflection} describe the
leading change in the bulk dispersion and momentum-to-angle map.
Both the lattice and quadratic models contain this leading
warping term. Their different reflection curves also involve
higher-order band corrections and the superconducting boundary
modes. In particular, the auxiliary roots of the quadratic
boundary problem lie outside the controlled valley expansion.
The present comparison therefore does not isolate the part of
the peak shift caused by bulk trigonal warping alone.

The maxima were refined in conserved transverse momentum rather than
read from the plotted grid. Starting grids of 201, 401, and 801 points
gave a largest peak-angle spread of $4.50\times10^{-6}$ degrees.
An independent Kwant scattering calculation \cite{Groth2014}, using
the same Hamiltonian but no analytical matching formulas, agrees with
the lattice result over 242 sampled incident states. The RMS and
maximum absolute differences in $R_A$ are $1.29\times10^{-14}$ and
$1.44\times10^{-13}$, respectively (Table~\ref{tab:checks}).

Lowering $E$ alone does not remove the disagreement. At $E/t_0=0.01$,
the superconducting chemical potential is still $\mu_S/t_0=1$, so the
S-side modes remain outside a controlled valley expansion. This is why
the next comparison reduces all junction energies together.

\begin{figure*}[!t]
\centering\includegraphics[width=\textwidth]{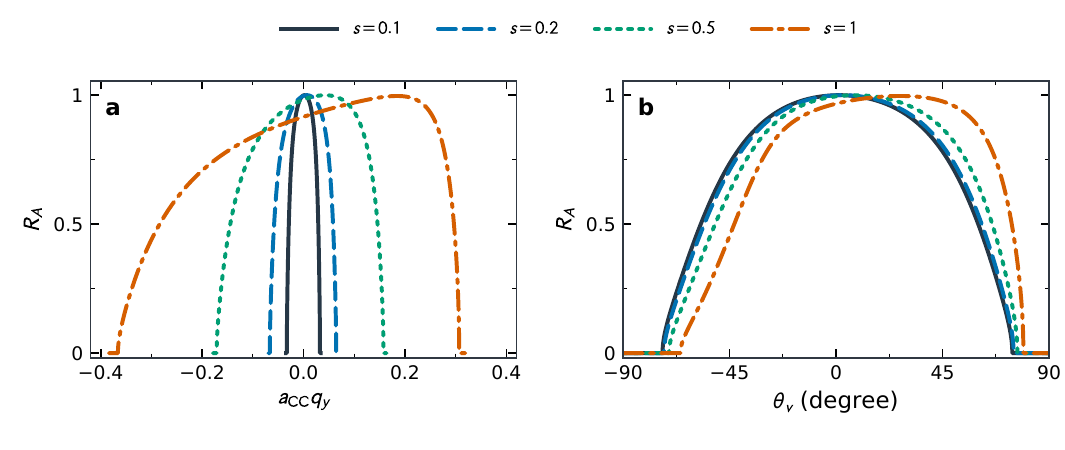}
\caption{Lattice Andreev reflection under the common scaling
$(\mu_N,\mu_S,\Delta_0,E)/t_0=s(0.5,1,0.1,0.01)$, with $\xi=+1$.
(a) Reflection probability versus the dimensionless Cartesian
transverse momentum $a_{\rm CC}q_y=p_Y/\sqrt3$.
(b) The same probabilities versus the Cartesian group-velocity
angle $\theta_v$. Each curve is drawn only where an incident
electron channel exists. The wave-vector angle used in
Ref.~\cite{ZhangYang2018} differs from $\theta_v$ in a warped band.
Lines connect the calculated points without smoothing.}
\label{fig:momentumangle}
\end{figure*}

\begin{figure*}[!t]
\centering\includegraphics[width=\textwidth]{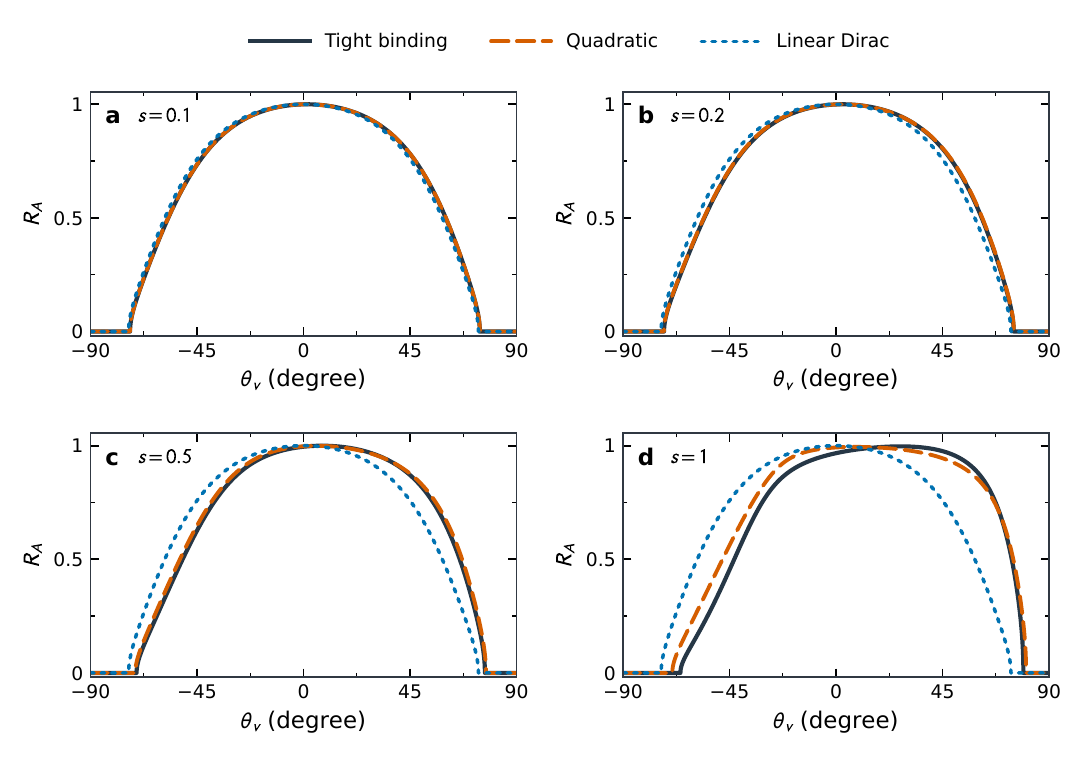}
\caption{Angular Andreev reflection under the common scaling in
Eq.~\eqref{eq:scaling}, with $E/\Delta_0=0.1$ and $\xi=+1$.
The panels show (a) $s=0.1$, (b) $0.2$, (c) $0.5$, and (d) $1$
on identical axes. Charcoal solid, orange dashed, and blue dotted
curves denote the tight-binding, quadratic, and linear-Dirac models.
Each model uses its own Cartesian group velocity to determine
$\theta_v$. The comparison includes the stated interface matching
and does not isolate a bulk warping correction.
Lines connect the calculated points without smoothing;
Fig.~S2 magnifies the central angular response.}
\label{fig:limit}
\end{figure*}

\subsection{Recovery of the low-energy limit}
We reduce all junction energies together according to
\begin{equation}
 (\mu_N,\mu_S,\Delta_0,E)=s(0.5,1,0.1,0.01)t_0,
 \label{eq:scaling}
\end{equation}
keeping $E/\Delta_0$, $\mu_N/\Delta_0$, and $\mu_S/\Delta_0$
fixed. This is a common energy scaling, rather than a change
of doping alone. Figure~\ref{fig:momentumangle} shows the same
lattice probabilities against transverse momentum and physical
incidence angle. The incident momentum interval narrows as $s$
decreases, while the angular response remains broad and becomes
more symmetric. Plotting both variables separates the shrinking
momentum range from the change in angular response.

Figure~\ref{fig:limit} compares all three models at the four scales.
The largest asymmetry and model differences occur at $s=1$;
the curves become much closer as the energies decrease.
At smaller physical $Q$, the trigonal-warping correction in
Eq.~\eqref{eq:warpedband} is weaker relative to the linear band.
The lattice and quadratic peak angles
decrease from $27.26^\circ$ and $8.38^\circ$ at $s=1$ to
$1.354^\circ$ and $1.285^\circ$ at $s=0.1$.
At fixed ratios, the linear-Dirac angular response is independent
of $s$. The finite-angle maxima are not claims of perfect
Andreev reflection; their heights remain below one for the
finite-doping example discussed above.

We quantify the agreement on 801 common physical angles in
$[-40^\circ,40^\circ]$. For a model $M$, define
\begin{equation}
 \varepsilon_M=\left[\frac1{801}\sum_{j=1}^{801}
 \bigl(R_A^M(\theta_j)-R_A^{\rm TB}(\theta_j)\bigr)^2\right]^{1/2},
\end{equation}
where $\theta_j=-40^\circ+80^\circ(j-1)/800$.
Each value is obtained by finding the transverse momentum
for the requested angle in each model and solving the scattering
problem directly. This interval avoids thresholds where an
open channel exists in only one model. As $s$ decreases from
1 to 0.1, the quadratic--lattice RMS difference falls from
$4.25\times10^{-2}$ to $3.04\times10^{-4}$, while the
Dirac--lattice difference falls from $9.82\times10^{-2}$ to
$8.67\times10^{-3}$. The intermediate scales $s=0.5$ and
0.2 follow the same decreasing trend. The quadratic model
is closer to the lattice throughout this tested sequence.
These four scales do not establish a universal convergence law.

Common scaling brings the physical modes in both regions
toward their valleys. Lowering $E$ alone at fixed doping
leaves their distance from the Dirac points set by $\mu_N$
and $\mu_S$. The comparison therefore supports checking the
energy scales on both sides of the junction. The quadratic
model still retains short-wavelength auxiliary modes, so
the observed convergence applies to the stated boundary
prescription and does not separate interface effects from
bulk-dispersion corrections.

\begin{figure*}[!t]
\centering\includegraphics[width=\textwidth]{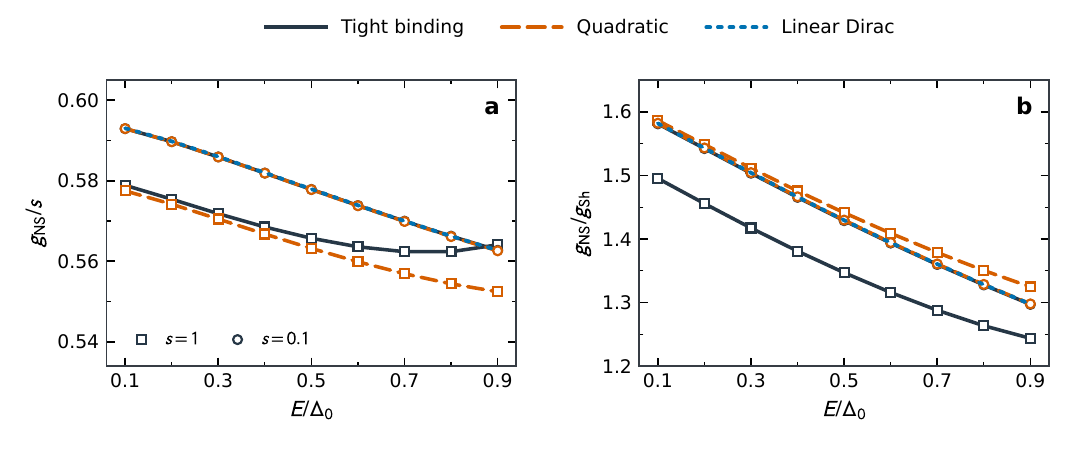}
\caption{Subgap conductance at
$(\mu_N,\mu_S,\Delta_0)/t_0=s(0.5,1,0.1)$, including both valleys.
(a) Conductance per common scale, $g_{NS}/s$.
(b) Conductance divided by each model's incident-mode reference
$g_{\rm Sh}$. Charcoal solid, orange dashed, and blue dotted curves
denote the tight-binding, quadratic, and linear-Dirac models.
Squares and circles distinguish $s=1$ and 0.1 for the lattice
and quadratic results. The two Dirac curves coincide in each
normalization and are drawn once. Every model uses the same
finite superconducting chemical potential, $\mu_S=2\mu_N$.
All nine bias points are direct conductance integrals; lines
connect them without smoothing.}
\label{fig:g}
\end{figure*}
\subsection{Differential conductance}

At zero temperature, the differential conductance is
$G_{NS}(V)=dI/dV$, evaluated at $E=eV>0$. Including spin and both
valleys, we write it per transverse repeat as
\begin{equation}
 g_{NS}\equiv\frac{G_{NS}}{(2e^2/h)N_Y}
 =\frac{1}{\pi}\sum_{\xi=\pm1}\int_{\mathcal I_\xi}dk_Y\,R_A.
 \label{eq:g}
\end{equation}
Here $\mathcal I_\xi$ is the interval of right-moving incident
electron modes. This follows from $1-R_N+R_A=2R_A$ in the
subgap Landauer--BTK expression
\cite{BlonderTinkhamKlapwijk1982,Beenakker1992,BhattacharjeeSengupta2006}.
The electron--hole scattering formalism keeps the charge signs of
the two outgoing species explicit \cite{AnantramDatta1996}.
For $N_Y$ transverse repeats with periodic boundary conditions,
adjacent phases differ by $2\pi/N_Y$. Thus
$N_Y^{-1}\sum_{k_Y}$ becomes $(2\pi)^{-1}\int dk_Y$.
The prefactor $2e^2/h$ contains spin degeneracy; the explicit
valley sum accounts for the two valleys. The other factor
of two follows from Andreev reflection, not another
degeneracy. Nambu electron and hole components must not be
counted as two independent incident species.
Integration in physical angle would require the Jacobian
$|dk_Y/d\theta_v|$; equally spaced angles do not count equal numbers
of transverse modes.

The endpoints are found from the normal-state dispersion. A closed
hole channel contributes zero to $R_A$, although its decaying mode
remains in the boundary solution. In the quadratic model we retain
the normal-state pocket connected to the valley centre; additional
zeros of the polynomial away from the valley are excluded from the
incident-channel count. The supplement gives the branch restriction,
analytic lattice endpoints, and quadrature checks.

Reducing $s$ reduces the number of incident channels. To distinguish
that change from a change in reflection probability, we separate mode
counting from scattering, as in transmission-channel treatments of
graphene transport \cite{Tworzydlo2006}. Define
\begin{align}
 g_{\rm Sh}&=\frac{1}{2\pi}\sum_\xi|\mathcal I_\xi|,\qquad
 \bar g_{NS}=\frac{g_{NS}}{g_{\rm Sh}},\notag\\
 \delta_g&=\frac{g_{NS}^{\rm QC}-g_{NS}^{\rm TB}}{g_{NS}^{\rm TB}}.
 \label{eq:normalization}
\end{align}
The Sharvin reference counts incident modes; it is not the conductance
of a normal junction with the same doping mismatch. Perfect Andreev
reflection in every incident mode gives $\bar g_{NS}=2$.

The linear-Dirac comparison uses wave-function continuity and the
same finite $\mu_S$ as the other models. Its incident interval is
$|p_Y|<2(\mu_N+E)/\sqrt3$, giving
\begin{equation}
 g_{\rm Sh}^{\rm D}=\frac{4(\mu_N+E)}{\pi\sqrt3}.
 \label{eq:diracsharvin}
\end{equation}
Using $p_Y=2(\mu_N+E)\sin\theta_v/\sqrt3$, the normalized
conductance can also be written as
\begin{align}
 \bar g_{NS}^{\rm D}&=2\int_0^{\theta_c}d\theta_v\,
 R_A^{\rm D}(\theta_v)\cos\theta_v,\notag\\
 \theta_c&=\arcsin\frac{\mu_N-E}{\mu_N+E}.
 \label{eq:diracangularconductance}
\end{align}
This is the ballistic normalization used in
Ref.~\cite{Beenakker2006}. Its strongly doped superconducting limit
differs from the finite ratio $\mu_S/\mu_N=2$ used here.
The bias dependence need not match those limiting curves.
In particular, a decreasing conductance here does not imply
specular Andreev reflection: $E<\mu_N$ throughout the plotted range.
At fixed energy ratios, the linear-Dirac reflection probabilities
are independent of $s$, while both $g_{NS}^{\rm D}$ and
$g_{\rm Sh}^{\rm D}$ scale with $s$. Its two scaled curves therefore
coincide in each panel of Fig.~\ref{fig:g}.

Over nine bias points, the largest absolute quadratic--lattice
conductance difference decreases from
$1.17\times10^{-2}$ at $s=1$ to $1.25\times10^{-5}$ at $s=0.1$.
The largest relative difference decreases from $2.07\%$ to
$0.0222\%$. For the linear-Dirac model, the largest relative
difference from the lattice decreases from $2.50\%$ to $0.0149\%$.
The channel-normalized comparison shows that convergence is not
merely a consequence of the shrinking conductance.

The two normalizations can give different orderings. At
$s=1$ and $E/\Delta_0=0.5$, the lattice and quadratic
conductances are 0.56571 and 0.56319. Their incident-mode
references are 0.42002 and 0.39077, giving normalized values
1.34686 and 1.44124. The quadratic model has a slightly
smaller absolute conductance but a larger value per incident
channel. This follows from its smaller channel count.

Across the nine biases, the largest absolute quadratic--lattice difference in
$\bar g_{NS}$ falls from $9.47\times10^{-2}$ to
$9.24\times10^{-4}$ as $s$ falls from 1 to 0.1.
The corresponding Dirac--lattice difference falls from
$8.73\times10^{-2}$ to $6.94\times10^{-4}$.
Thus the quadratic model's smaller central angular error does
not guarantee a smaller conductance error. Integration averages
the angular response, while normalization also changes with the
incident-channel count. Angular reflection and conductance provide
complementary tests of the continuum approximation.

\begin{table}[!t]
\caption{Independent numerical checks. Differences refer to
$R_A$ unless stated otherwise. The fresh conductance checks
cover two models, two scales, and three biases; each value
includes both valleys.}
\label{tab:checks}
\centering
\begin{tabular}{@{}lrr@{}}
\toprule
Comparison & Samples & Largest difference\\
\midrule
Lattice--Kwant, original & 242 & $1.44\times10^{-13}$\\
Lattice--Kwant, fresh & 80 & $3.38\times10^{-12}$\\
Four-row--augmented, fresh & 80 & $2.66\times10^{-15}$\\
Adaptive--stored $g_{NS}$ & 12 & $2.35\times10^{-14}$\\
\bottomrule
\end{tabular}
\end{table}
\subsection{Numerical checks}
The checks separate the accuracy of a solver from the accuracy
of a model approximation. Residuals and agreement with Kwant
test the stated lattice boundary problem. Differences between
the lattice and continuum observables test the approximation.
We check the Bloch, eigenvector, boundary, and flux residuals.
The largest analytical flux error is
$6.32\times10^{-11}$ and largest normalized boundary residual is
$1.53\times10^{-16}$. The Kwant unitarity error is below
$1.74\times10^{-14}$. These checks establish numerical consistency
for the stated boundary models. A separate comparison over
254 channels gives a largest finite-mode coefficient difference
of $1.46\times10^{-15}$ between the augmented lattice system
and its four-equation reduction. The largest extra boundary
coefficient is $1.13\times10^{-15}$, consistent with
Eq.~\eqref{eq:vanishing}.

Conductance is integrated only after locating the electron and
hole channel endpoints. A sine change of variable clusters the
quadrature nodes near channel closure. Increasing the order from
160 to 320 and independently widening the quadratic search
interval change the valley contributions by less than
$10^{-13}$ in the supplied checks. These are observed
repeatability differences, not rigorous error bounds.
The supplement gives the residual definitions and full tables.

We also check a fresh set of 80 incident states spanning both
valleys, $s=1,0.5,0.2,0.1$, and $E/\Delta_0=0.1,0.7$.
Five transverse positions per case include points close to the
edges of the incident interval, as well as closed hole channels.
The largest lattice--Kwant difference in this set is
$3.38\times10^{-12}$. The larger value than in the original
242-state comparison reflects the different sample set; the
quoted numerical accuracy is always tied to its tested states.
Twelve separate adaptive integrations, performed directly in
transverse momentum without the sine substitution, reproduce
the stored conductances to $2.35\times10^{-14}$. Table~\ref{tab:checks}
summarizes these checks. Symbolic checks verify the determinant
and auxiliary roots.

The added linear-Dirac conductance is also checked using an
independent Cartesian-basis mode solve. Its reflection probabilities
agree with the tilted-coordinate implementation to
$2.33\times10^{-15}$ over 84 tested channels. Six adaptive
integrals agree with the sine-substitution quadrature to
$1.90\times10^{-14}$, and increasing the quadrature order from
128 to 256 changes the Dirac conductance by at most
$2.18\times10^{-14}$. Increasing $\mu_S$ at fixed normal-region
parameters also recovers the strongly doped superconducting
limit of Ref.~\cite{Beenakker2006}.

\subsection{Physical range and experimental relevance}
Continuum accuracy requires a separate small-parameter condition. Restoring dimensional energies,
\begin{equation}
 \max\bigl(|\mu_N+E|,|\mu_N-E|,|\mu_S+\ii\Omega|,
 |\mu_S-\ii\Omega|\bigr)\ll t_0.
 \label{eq:validity}
\end{equation}
The physical roots must also satisfy $\max(|p_X|,|p_Y|)\ll1$.
Neither condition turns the auxiliary roots in Eq.~\eqref{eq:auxiliary}
into microscopic lattice modes. The convergence observed here therefore
belongs to the specified sharp-interface prescription and parameter
sequence. It does not isolate a bulk-dispersion correction or establish
a pseudospin criterion for perfect Andreev reflection. The large
parameter set is a test of the approximation, not an estimate of an
induced gap in a particular device.
In particular, $\mu_S=t_0$ reaches the normal-state van Hove
energy of the nearest-neighbour model. A small excitation
energy does not make this S-side band a Dirac dispersion.
Exact lattice scattering here means exact scattering within
the specified nearest-neighbour, prescribed-gap Hamiltonian.

The experimental relevance is methodological. Measured
conductance also depends on contact doping, transparency,
geometry, and temperature. Bilayer graphene--NbSe$_2$
experiments have probed the retro-to-specular crossover
\cite{Efetov2016}; their bilayer dispersion differs from the
monolayer model here. They motivate careful band and channel
identification, rather than a direct quantitative comparison.

A device calculation would require measured or independently
constrained chemical potentials and induced gap. Finite
contact length can introduce repeated reflections, whereas
S is a half-space here. Temperature averages a spectrum over
an energy window of order $k_BT$. These effects are not in the
displayed results. The present benchmark identifies when a
continuum interface should be checked against the lattice
before fitted parameters are given a microscopic interpretation.
\section{Conclusions}

For a graphene NS step with unchanged hopping across the interface, the tight-binding equations yield four independent matching conditions, and the resulting reflection probabilities agree with an independent Kwant calculation for the channels tested. The quadratic continuum model can predict a substantially different angular position of the reflection maximum when the junction energies lie outside the controlled low-energy regime. Reducing the chemical potentials, superconducting gap, and excitation energy together at fixed ratios brings the central angular response and channel-normalized conductance closer to the lattice results. The leading trigonal-warping correction, captured by the quadratic valley expansion, modifies the constant-energy contours and group-velocity directions relative to the linear Dirac description. However, the discrepancies between the lattice and continuum results cannot be attributed to this correction alone. Assessing the validity of a continuum description requires consideration of the relevant energy scales and modes in both the normal and superconducting regions.

An important unresolved issue concerns the interface boundary conditions. The quadratic Hamiltonian with sharp-step derivative matching defines a particular continuum model, but its auxiliary modes lie outside the controlled range of the valley expansion. A continuum boundary condition derived directly from the lattice model, or a comparison with a controlled smooth-interface treatment, would help distinguish boundary-related effects from bulk band-structure corrections. For terminated graphene, the relationship between microscopic edge structure and Dirac boundary conditions has been derived explicitly \cite{AkhmerovBeenakker2008Boundary}. A corresponding derivation for the graphene NS interface considered here would provide a useful basis for resolving this remaining question.
\appendix
\section{Transverse-channel intervals}
The conductance measure requires the projection of the incident
contour onto the interface. In the positive-valley patch
$0<k_Y<\pi$, set $q=2\cos(k_Y/2)>0$. Then
$\eta=q e^{\ii k_Y/2}$ and
\begin{equation}
 |f|^2=1+q^2+2q\cos(k_X+k_Y/2).
\end{equation}
For real $k_X$, this spans $[(1-q)^2,(1+q)^2]$.
For $0<|\Lambda|<1$, a propagating state exists precisely
when $1-|\Lambda|\leq q\leq1+|\Lambda|$. Since the cosine
decreases on this patch, the endpoints in $p_Y=k_Y-K_0$ are
\begin{align}
 \ell_+(\Lambda)&=2\arccos[(1+|\Lambda|)/2]-K_0,\notag\\
 r_+(\Lambda)&=2\arccos[(1-|\Lambda|)/2]-K_0.
\end{align}
The other valley has interval $[-r_+,-\ell_+]$.
Electron and hole intervals use $\Lambda=\mu_N+E$
and $\mu_N-E$. Only their intersection carries nonzero
Andreev flux. The other incident electron modes still
count in the Sharvin reference.

For the quadratic model the endpoints satisfy
\begin{equation}
 F_\xi(p_X,p_Y)=\Lambda^2,\qquad
 \partial_{p_X}F_\xi(p_X,p_Y)=0.
\end{equation}
The repeated root has zero longitudinal velocity. The
incident pocket is followed from the valley origin;
distant zeros of the truncated polynomial are excluded
from the incoming channel count. This is separate from
retaining auxiliary evanescent modes in the boundary solve.

On an integration interval $[\ell,r]$, use
\begin{equation}
 p_Y=\frac{\ell+r}{2}+\frac{r-\ell}{2}\sin u,\qquad
 -\frac\pi2<u<\frac\pi2.
\end{equation}
The Jacobian is $(r-\ell)\cos u/2$. Gauss--Legendre
nodes in $u$ avoid the zero-current endpoints. Each
node is solved directly from the matching equations;
the integration does not interpolate an angular plot.

\section{Polynomial used in the quadratic solve}
For reproducibility, we give the longitudinal polynomial of
the second-order Hamiltonian at fixed real $p_Y$. Set $a_\xi=e^{-\ii\xi K_0}$ and
$b_\xi=e^{\ii\xi K_0}$. Alongside Eq.~\eqref{eq:fquad},
the analytically continued partner is
\begin{equation}
 \bar f_\xi=-\ii b_\xi p_X+\ii a_\xi p_Y
 +\tfrac12(b_\xi p_X^2+a_\xi p_Y^2).
\end{equation}
This expression conjugates the coefficients on the real
momentum axis and then continues $p_X$ as an independent
complex variable. The products satisfy
$a_\xi b_\xi=1$, $a_\xi^2+b_\xi^2=-1$, and
$a_\xi^2-b_\xi^2=\ii\xi\sqrt3$. Multiplying the two
matrix entries therefore gives
\begin{align}
 F_\xi={}&p_X^2+p_Xp_Y+p_Y^2
 -\frac{\xi\sqrt3}{2}p_Xp_Y(p_X+p_Y)\notag\\
 &+\frac14(p_X^4+p_Y^4-p_X^2p_Y^2).
 \label{eq:Fexpanded}
\end{align}
Thus each target $\Lambda$ is obtained from
\begin{equation}
 \frac14p_X^4+c_2p_X^2+c_1p_X+c_0=0,
\end{equation}
where
\begin{align}
 c_2&=1-\frac{\xi\sqrt3}{2}p_Y-\frac14p_Y^2,\notag\\
 c_1&=p_Y-\frac{\xi\sqrt3}{2}p_Y^2,\notag\\
 c_0&=p_Y^2+\frac14p_Y^4-\Lambda^2.
\end{align}
There is no cubic term. The coefficients $c_1,c_2$ are real,
but $c_0$ is complex for a subgap superconducting target.
Two quartic solves, one for each target, give the regional modes.
Sorting roots by their numerical order has no physical
meaning; propagation current or decay direction selects
the retained columns.

The fourth-order terms in $F_\xi$ arise by multiplying
two second-order matrix entries. They do not make the
Hamiltonian a fourth-order expansion of the lattice
Hamiltonian. In particular, omitted cubic terms in the
lattice structure factor would also contribute to the
fourth-order dispersion. The polynomial defines the
quadratic model used here exactly, within its stated
long-wavelength scope.

% Slightly tighter reference leading keeps the complete bibliography on page 11.
\renewcommand{\bibfont}{\small\linespread{0.97}\selectfont}
\bibliography{references}
\end{document}